\documentclass[a4paper,draftcls,onecolumn,technote,11pt]{IEEEtran}%
\usepackage{amsfonts}
\usepackage{amsmath}
\usepackage{amssymb}
\usepackage{graphicx}%
\providecommand{\U}[1]{\protect\rule{.1in}{.1in}}
\begin{document}

\title{Additive-State-Decomposition Dynamic Inversion Stabilized Control for a Class
of Uncertain MIMO Systems}
\author{Quan Quan\thanks{Corresponding Author: Quan Quan, Associate Professor,
Department of Automatic Control, Beijing University of Aeronautics and
Astronautics, Beijing 100191, qq\_buaa@buaa.edu.cn,
http://quanquan.buaa.edu.cn.}, Guangxun Du and Kai-Yuan Cai}
\maketitle

\begin{abstract}
This paper presents a new control, namely additive-state-decomposition dynamic
inversion stabilized control, that is used to stabilize a class of multi-input
multi-output (MIMO) systems subject to nonparametric time-varying
uncertainties with respect to both state and input. By additive state
decomposition and a new definition of output, the considered uncertain system
is transformed into a minimum-phase uncertainty-free system with relative
degree one, in which all uncertainties are lumped into a new disturbance at
the output. Subsequently, dynamic inversion control is applied to reject the
lumped disturbance. Performance analysis of the resulting closed-loop dynamics
shows that the stability can be ensured. Finally, to demonstrate its
effectiveness, the proposed control is applied to two existing problems by
numerical simulation. Furthermore, in order to show its practicability, the
proposed control is also performed on a real quadrotor to stabilize its
attitude when its inertia moment matrix is subject to a large uncertainty.

\end{abstract}

\section{Introduction}

Stabilization in control systems with uncertainties depending on state and
input has attracted the interest of many researchers. Uncertainties depending
on state originate from various sources, including variations in plant
parameters and inaccuracies that arise from identification. Input
uncertainties include uncertain gains, dead zone nonlinearities, quantization,
and backlash. In practice, these uncertainties may degrade or destabilize
system performance. For example, given that aerodynamic parameters are
functions of flight conditions, some aircraft are nonlinear and undergo rapid
parameter variations. These attributes stem from the fact that aircraft can
operate in a wide range of aerodynamic conditions. As a result, aerodynamic
parameters, which exist in system and input matrices \cite{Apkarian(1995)}%
,\cite{Malloy(1998)},\cite{Young(2007)}, are inherently uncertain. Therefore,
robust stabilization control problems for systems with uncertainties depending
on both state and input are important.

In this paper, a stabilization control problem is investigated for a class of
multi-input multi-output (MIMO) systems subject to nonparametric time-varying
uncertainties with respect to both state and input. Several accepted control
methods for handling uncertainties are briefly reviewed. A direct approach is
to estimate all unknown parameters, and then simultaneously use such
parameters to resolve uncertainties. Lyapunov methods are adopted in analyzing
the stability of closed-loop systems. In \cite{Young(2007)}, nonparametric
uncertainties involving state and input are approximated via basis functions
with unknown parameters, which are estimated by given adaptive laws. With the
estimated parameters, an approximate dynamic inversion method was proposed. It
is in fact an adaptive dynamic inversion method \cite{Hovakimyan(2008)}%
,\cite{Lavretsky(2008)}. In \cite{Cao(2010)}, $\mathcal{L}_{1}$ adaptive
control architecture was proposed for systems with an unknown input gain, as
well as unknown time-varying parameters and disturbances. In
\cite{Nalin(2006)}, adaptive feedback control was used to track the desired
angular velocity trajectory of a planar rigid body with unknown rotational
inertia and unknown input nonlinearity. In \cite{MacKunis(2010)}, two
asymptotic tracking controllers were designed for the output tracking of an
aircraft system under parametric uncertainties and unknown nonlinear
disturbances, which are not linearly parameterizable. An adaptive extension
was then presented, in which the feedforward adaptive estimates of input
uncertainties are used. As indicated in \cite{Young(2007)}%
-\cite{MacKunis(2010)}, adaptive controllers may require numerous integrators
that correspond to unknown parameters in an uncertain system. Each unknown
parameter requires an integrator for estimation, thereby resulting in a
closed-loop system with a reduced stability margin. In addition, the estimates
may not approach real parameters without the persistent excitation of signals,
which are difficult to generate in practice, particularly under numerous
unknown parameters \cite{Landau(2011)}. The second direct method for resolving
uncertainties is designing inverse control by a neural network that cancels
input nonlinearities, thereby generating a linear function
\cite{Singhal(2009)},\cite{Chen(2010)}. In contrast to traditional inverse
control schemes, a neural network approximates an unknown nonlinear term
\cite{Zhang(2003)}. Neural network methods can also be considered as adaptive
control methods, except that they have different basis functions. Thus, they
also have the same problems as those encountered in adaptive control methods.
The third approach is adopting sliding mode control, which presents inherent
fast response and insensitivity to plant parameter variation and/or external
perturbation. In \cite{Hsu(1998)}, a new sliding mode control law based on the
measurability of all system states was presented. The law ensures global reach
conditions of the sliding mode for systems subject to nonparametric
time-varying uncertainties with respect to both state and input. Along this
idea of \cite{Hsu(1998)}, an output feedback controller was further proposed
in \cite{Shen(1999)}. Sliding mode controllers essentially rely on infinite
gains to achieve good tracking performance, which is not always feasible in
practice. In practice, moreover, switching will consume energy and may excite
high-frequency modes.

To overcome these drawbacks, this paper proposes a stabilization approach that
involves dynamic inversion based on additive state decomposition (ASD).
Additive state decomposition \cite{Quan(2009)} is different from the
lower-order subsystem decomposition methods existing in the literature.
Concretely, taking the system $\dot{x}\left(  t\right)  =f\left(  t,x\right)
,x\in%
%TCIMACRO{\U{211d} }%
%BeginExpansion
\mathbb{R}
%EndExpansion
^{n}$ for example, it is decomposed into two subsystems: $\dot{x}_{1}\left(
t\right)  =f_{1}\left(  t,x_{1},x_{2}\right)  $ and $\dot{x}_{2}\left(
t\right)  =f_{2}\left(  t,x_{1},x_{2}\right)  $, where $x_{1}\in%
%TCIMACRO{\U{211d} }%
%BeginExpansion
\mathbb{R}
%EndExpansion
^{n_{1}}\ $and$\ x_{2}\in%
%TCIMACRO{\U{211d} }%
%BeginExpansion
\mathbb{R}
%EndExpansion
^{n_{2}},$ respectively. The lower-order subsystem decomposition satisfies
$n=n_{1}+n_{2}$ and $x=x_{1}\oplus x_{2}.$ By contrast, the proposed additive
state decomposition satisfies $n=n_{1}=n_{2}$ and$\ x=x_{1}+x_{2}.$ The key
idea of the proposed method is that it combines nonparametric time-varying
uncertainties with respect to both state and input into one disturbance by
ASD. Such a disturbance is then compensated for. The proposed controller is
continuous and enables asymptotic stability in the presence of time-invariant
uncertainties. Moreover, by choosing a special filter, the proposed
controllers can be finally replaced by proportional-integral (PI) controllers.
This is consistent with the controller form in \cite{Teo(2010)} for a similar
problem. However, compared with \cite{Teo(2010)}, the considered plant,
analysis method and design procedure are all different, especially the
analysis method and design procedure. The bound on a parameter, corresponding
to the singular perturbation parameter in \cite{Teo(2010)}, is also given
explicitly. Moreover, our proposed controllers are not just in the form of PI
controllers. This paper focuses on a stabilization problem, which
distinguishes it from the authors' previous work on ASD. In previous research,
ASD has been applied to tracking problems for nonlinear systems without
stabilization problems or with the stabilization problem being solved by a
simple state feedback controller \cite{Quan(2013)},\cite{Quan(IJRNC)}%
,\cite{Quan(IJSS)}. The other additive decomposition, namely additive output
decomposition \cite{Quan(CDC2012)}, is also applied to a tracking problem with
a stable controlled plant.

\section{Problem Formulation and Additive State Decomposition}

\subsection{Problem Formulation}

Consider a class of MIMO systems subject to nonparametric time-varying
uncertainties with respect to both state and input as follows:%
\begin{equation}
\dot{x}\left(  t\right)  =A_{0}x\left(  t\right)  +B\left(  h\left(
t,u\right)  +\sigma\left(  t,x\right)  \right)  ,x\left(  0\right)  =x_{0}
\label{perturbedsystem}%
\end{equation}
where $x\left(  t\right)  \in%
%TCIMACRO{\U{211d} }%
%BeginExpansion
\mathbb{R}
%EndExpansion
^{n}$ is the system state (taken as a measurable output), $u\left(  t\right)
\in%
%TCIMACRO{\U{211d} }%
%BeginExpansion
\mathbb{R}
%EndExpansion
^{m}$\ is the control, $A_{0}\in%
%TCIMACRO{\U{211d} }%
%BeginExpansion
\mathbb{R}
%EndExpansion
^{n\times n}$ is a known matrix, $B\in%
%TCIMACRO{\U{211d} }%
%BeginExpansion
\mathbb{R}
%EndExpansion
^{n\times m}$ is a known constant matrix, $h:\left[  0,\infty\right)  \times%
%TCIMACRO{\U{211d} }%
%BeginExpansion
\mathbb{R}
%EndExpansion
^{m}\rightarrow%
%TCIMACRO{\U{211d} }%
%BeginExpansion
\mathbb{R}
%EndExpansion
^{m}$ is an unknown nonlinear vector function, and $\sigma:\left[
0,\infty\right)  \times%
%TCIMACRO{\U{211d} }%
%BeginExpansion
\mathbb{R}
%EndExpansion
^{n}\rightarrow%
%TCIMACRO{\U{211d} }%
%BeginExpansion
\mathbb{R}
%EndExpansion
^{m}$\ is an unknown nonlinear time-varying disturbance. For system
(\ref{perturbedsystem}), the following assumptions are made.

\textbf{Assumption 1}. The pair $\left(  A_{0},B\right)  $ is controllable.

\textbf{Assumption 2}. The unknown nonlinear vector function$\ h$ satisfies
$h\left(  t,0\right)  \equiv0,\ \left\Vert \frac{\partial h}{\partial
t}\right\Vert \leq l_{h_{t}}\left\Vert u\right\Vert ,$ $\frac{\partial
h}{\partial u}>\underline{l}_{h_{u}}I_{m}$ and $\left\Vert \frac{\partial
h}{\partial u}\right\Vert \leq\overline{l}_{h_{u}},$ $\forall u\in%
%TCIMACRO{\U{211d} }%
%BeginExpansion
\mathbb{R}
%EndExpansion
^{m},$ $\forall t\geq0,$ where $l_{h_{t}},\underline{l}_{h_{u}},\overline
{l}_{h_{u}}>0.$

\textbf{Assumption 3}. The time-varying disturbance$\ \sigma\left(
t,x\right)  $ satisfies $\left\Vert \sigma\left(  t,x\right)  \right\Vert \leq
k_{\sigma}\left\Vert x\left(  t\right)  \right\Vert +\delta_{\sigma}\left(
t\right)  ,\ \left\Vert \frac{\partial\sigma}{\partial x}\right\Vert \leq
l_{\sigma_{x}}$ and $\left\Vert \frac{\partial\sigma}{\partial t}\right\Vert
\leq l_{\sigma_{t}}\left\Vert x\left(  t\right)  \right\Vert +d_{\sigma
}\left(  t\right)  ,$ where $\frac{\partial\sigma}{\partial x}\in%
%TCIMACRO{\U{211d} }%
%BeginExpansion
\mathbb{R}
%EndExpansion
^{m\times n},$ $k_{\sigma},\delta_{\sigma}\left(  t\right)  ,l_{\sigma_{x}%
},l_{\sigma_{t}},d_{\sigma}\left(  t\right)  >0,$ $\forall t\geq0$, and
$\delta_{\sigma}\left(  t\right)  ,d_{\sigma}\left(  t\right)  $ are bounded.

\textbf{Remark 1}. If $h$ is a dead zone function such as $h\left(
t,u\right)  =\left\{
\begin{array}
[c]{c}%
u\\
0
\end{array}
\right.
\begin{array}
[c]{c}%
\left\vert u_{i}\right\vert \geq\mu,i=1,\cdots,m\\
\left\vert u_{i}\right\vert <\mu,i=1,\cdots,m
\end{array}
$,\ then it can be reformulated as $h\left(  t,u\right)  =u+\delta\left(
t\right)  ,$ where $\left\Vert \delta\left(  t\right)  \right\Vert \leq\mu.$
In practice, the parameters $l_{h_{t}},\underline{l}_{h_{u}},\overline
{l}_{h_{u}},k_{\sigma},\delta_{\sigma},l_{\sigma_{x}},l_{\sigma_{t}}%
,d_{\sigma}\ $need not be known.

The control objective is to design a stabilized controller to drive the system
state such that $x\left(  t\right)  \rightarrow0$ as $t\rightarrow\infty$ or
the state is ultimately bounded by a small value. In the following, for
convenience, the notation $t$ will be dropped except\ when necessary for clarity.

\subsection{Additive State Decomposition}

In order to make the paper self-contained, ASD in \cite{Quan(2009)}%
,\cite{Quan(2013)},\cite{Quan(IJRNC)},\cite{Quan(IJSS)} is recalled briefly
here. Consider the following `original' system:%
\begin{equation}
f\left(  {t,\dot{x},x}\right)  =0,x\left(  0\right)  =x_{0}
\label{Gen_Orig_Sys}%
\end{equation}
where $x\in%
%TCIMACRO{\U{211d} }%
%BeginExpansion
\mathbb{R}
%EndExpansion
^{n}$. First, a `primary' system is brought in, having the same dimension as
(\ref{Gen_Orig_Sys}):%
\begin{equation}
f_{p}\left(  {t,\dot{x}_{p},x_{p}}\right)  =0,x{_{p}}\left(  0\right)
=x_{p,0} \label{Gen_Pri_Sys}%
\end{equation}
where ${x_{p}}\in%
%TCIMACRO{\U{211d} }%
%BeginExpansion
\mathbb{R}
%EndExpansion
^{n}$. From the original system (\ref{Gen_Orig_Sys}) and the primary system
(\ref{Gen_Pri_Sys}), the following `secondary' system is derived:%
\begin{equation}
f\left(  {t,\dot{x},x}\right)  -f_{p}\left(  {t,\dot{x}_{p},x_{p}}\right)
=0,x\left(  0\right)  =x_{0} \label{Gen_Sec_Sys0}%
\end{equation}
where ${x_{p}}\in%
%TCIMACRO{\U{211d} }%
%BeginExpansion
\mathbb{R}
%EndExpansion
^{n}$ is given by the primary system (\ref{Gen_Pri_Sys}). A new variable
${x_{s}}\in%
%TCIMACRO{\U{211d} }%
%BeginExpansion
\mathbb{R}
%EndExpansion
^{n}$ is defined as follows:%
\begin{equation}
{x_{s}\triangleq x-x_{p}}. \label{Gen_RelationPS}%
\end{equation}
Then the secondary system (\ref{Gen_Sec_Sys0}) can be further written as
follows:%
\begin{equation}
f\left(  {t,\dot{x}_{s}+\dot{x}_{p},x_{s}+x_{p}}\right)  -f_{p}\left(
{t,\dot{x}_{p},x_{p}}\right)  =0,x{_{s}}\left(  0\right)  =x_{0}-x_{p,0}.
\label{Gen_Sec_Sys}%
\end{equation}
From the definition (\ref{Gen_RelationPS}), it follows%
\begin{equation}
{x}\left(  t\right)  ={x_{p}\left(  t\right)  +x_{s}\left(  t\right)
,t\geq0.} \label{Gen_RelationPS1}%
\end{equation}

\textbf{Remark 2}\textit{.} By ASD\textit{, }the\textit{\ }system
(\ref{Gen_Orig_Sys}) is decomposed into two subsystems with the same dimension
as the original system. In this sense our decomposition is \textquotedblleft
additive\textquotedblright. In addition, this decomposition is with respect to
state. So, it is called \textquotedblleft additive state
decomposition\textquotedblright\emph{.}

As a special case of (\ref{Gen_Orig_Sys}), a class of differential dynamic
systems is considered as follows:%
\begin{align}
\dot{x}  &  =f\left(  {t,x}\right)  ,x\left(  0\right)  =x_{0},\nonumber\\
y  &  =g\left(  {t,x}\right)  \label{Dif_Orig_Sys}%
\end{align}
where ${x}\in%
%TCIMACRO{\U{211d} }%
%BeginExpansion
\mathbb{R}
%EndExpansion
^{n}$ and $y\in%
%TCIMACRO{\U{211d} }%
%BeginExpansion
\mathbb{R}
%EndExpansion
^{m}.$ Two systems, denoted by the primary system and (derived) secondary
system respectively, are defined as follows:%
\begin{align}
\dot{x}_{p}  &  =f_{p}\left(  {t,x_{p}}\right)  ,x_{p}\left(  0\right)
=x_{p,0}\nonumber\\
y_{p}  &  =g_{p}\left(  {t,x}_{p}\right)  \label{Dif_Pri_Sys}%
\end{align}
and%
\begin{align}
\dot{x}_{s}  &  =f\left(  {t,x_{p}}+{x_{s}}\right)  -f_{p}\left(  {t,x_{p}%
}\right)  ,x_{s}\left(  0\right)  =x_{0}-x_{p,0},\nonumber\\
y_{s}  &  =g\left(  {t,x_{p}}+{x_{s}}\right)  -g_{p}\left(  {t,x}_{p}\right)
\label{Dif_Sec_Sys}%
\end{align}
where ${x_{s}}\triangleq{x-x_{p}}$ and $y_{s}\triangleq{y-y_{p}}$. The
secondary system (\ref{Dif_Sec_Sys}) is determined by the original system
(\ref{Dif_Orig_Sys}) and the primary system (\ref{Dif_Pri_Sys}). From the
definition, it follows%
\begin{equation}
{x}\left(  t\right)  ={x_{p}\left(  t\right)  +x_{s}\left(  t\right)
,y\left(  t\right)  =y_{p}\left(  t\right)  +y_{s}\left(  t\right)  ,t\geq0.}
\label{Gen_RelationDif}%
\end{equation}

\section{Additive-State-Decomposition Dynamic Inversion Stabilized Control}

In this section, by ASD, the considered uncertain system is first transformed
into an uncertainty-free system but subject to a lumped disturbance at the
output. Then a dynamic inversion method is applied to this transformed system.
Finally, the performance of the resultant closed-loop system is analyzed.

\subsection{Output Matrix Redefinition}

Since the pair $\left(  A_{0},B\right)  $ is controllable by
\textit{Assumption 1}, a vector $K\in%
%TCIMACRO{\U{211d} }%
%BeginExpansion
\mathbb{R}
%EndExpansion
^{n\times m}$ can be always found such that $A_{0}+BK^{T}$ is stable. This
also implies that there exist $0<P,M\in%
%TCIMACRO{\U{211d} }%
%BeginExpansion
\mathbb{R}
%EndExpansion
^{n\times n}$ such that%
\begin{equation}
P\left(  A_{0}+BK^{T}\right)  +\left(  A_{0}+BK^{T}\right)  ^{T}P=-M.
\label{PQ}%
\end{equation}
According to this, the system (\ref{perturbedsystem}) is rewritten to be%
\begin{equation}
\dot{x}=Ax+B\left(  h\left(  t,u\right)  -K^{T}x+\sigma\left(  t,x\right)
\right)  ,x\left(  0\right)  =x_{0} \label{perturbedsystem_tran}%
\end{equation}
where $A=A_{0}+BK^{T}.$ Based on matrix $A$, a new definition of output matrix
$C$ is given in the following theorem.

\textbf{Theorem 1. }Under \textit{Assumption 1},\textbf{ }suppose $A^{T}%
$\ has\textbf{ }$n$ negative real eigenvalues, denoted by $-\lambda_{i}\in%
%TCIMACRO{\U{211d} }%
%BeginExpansion
\mathbb{R}
%EndExpansion
,$ to which correspond $n$ independent unit real eigenvectors, denoted by
$c_{i}\in%
%TCIMACRO{\U{211d} }%
%BeginExpansion
\mathbb{R}
%EndExpansion
^{n},$ $i=1,\cdots,n.$ If the output matrix is proposed as%
\begin{equation}
C=\left[
\begin{array}
[c]{ccc}%
c_{1} & \cdots & c_{m}%
\end{array}
\right]  \in%
%TCIMACRO{\U{211d} }%
%BeginExpansion
\mathbb{R}
%EndExpansion
^{n\times m} \label{output matrix}%
\end{equation}
then%
\[
C^{T}A=-\Lambda C^{T}%
\]
where $\Lambda=$diag$\left(  \lambda_{1},\cdots,\lambda_{m}\right)  \in%
%TCIMACRO{\U{211d} }%
%BeginExpansion
\mathbb{R}
%EndExpansion
^{m\times m}.$ Furthermore, if $\Lambda$ has the form $\Lambda=\alpha I_{m},$
then det$\left(  C^{T}B\right)  \neq0.$

\textit{Proof}. Since $c_{i},$ $i=1,\cdots,m$ are $m$ independent unit
eigenvectors of $A^{T}$, it follows%
\[
A^{T}c_{i}=-\lambda_{i}c_{i},i=1,\cdots,m,
\]
namely $A^{T}C=-C\Lambda$. Then $C^{T}A=-\Lambda C^{T}.$ Next, det$\left(
C^{T}B\right)  \neq0\ $will be shown. Suppose, to the contrary, that
det$\left(  C^{T}B\right)  =0.$ According to this, there exists a nonzero
vector $w\in%
%TCIMACRO{\U{211d} }%
%BeginExpansion
\mathbb{R}
%EndExpansion
^{m}$ such that $w^{T}C^{T}B=0.$ Define $v=Cw.$ Since $w\neq0$ and $C\ $is of
column full rank, it follows $v\neq0.$ With such a vector $v,$ it further
follows%
\begin{align*}
v^{T}\left[
\begin{array}
[c]{cccc}%
B & AB & \cdots & A^{n-1}B
\end{array}
\right]   &  =w^{T}C^{T}\left[
\begin{array}
[c]{cccc}%
B & AB & \cdots & A^{n-1}B
\end{array}
\right] \\
&  =w^{T}\left[
\begin{array}
[c]{cccc}%
C^{T}B & C^{T}AB & \cdots & C^{T}A^{n-1}B
\end{array}
\right] \\
&  =w^{T}\left[
\begin{array}
[c]{cccc}%
C^{T}B & -\alpha C^{T}B & \cdots & \left(  -\alpha\right)  ^{n-1}C^{T}B
\end{array}
\right] \\
&  =w^{T}C^{T}B\left[
\begin{array}
[c]{cccc}%
I_{m} & -\alpha I_{m} & \cdots & \left(  -\alpha\right)  ^{n-1}I_{m}%
\end{array}
\right] \\
&  =0.
\end{align*}
This implies that rank$\left[
\begin{array}
[c]{cccc}%
B & AB & \cdots & A^{n-1}B
\end{array}
\right]  <n,$ namely $\left(  A,B\right)  $ is uncontrollable and $\left(
A_{0},B\right)  $ is further uncontrollable, which contradicts the assumption
that the pair $\left(  A_{0},B\right)  $ is controllable. Then det$\left(
C^{T}B\right)  \neq0.$ $\square$

By \textit{Theorem 1},\textbf{\ }a virtual output $y=C^{T}x\ $is defined,
whose first derivative is%
\begin{align}
\dot{y}  &  =C^{T}\dot{x}\nonumber\\
&  =C^{T}Ax+C^{T}B\left(  h\left(  t,u\right)  -K^{T}x+\sigma\left(
t,x\right)  \right) \nonumber\\
&  =-\Lambda y+C^{T}B\left(  h\left(  t,u\right)  -K^{T}x+\sigma\left(
t,x\right)  \right)  . \label{externalsystem}%
\end{align}
Then, the system (\ref{perturbedsystem_tran}) can be rewritten as%
\begin{equation}
\left[
\begin{array}
[c]{c}%
\dot{\eta}\\
\dot{y}%
\end{array}
\right]  =\left[
\begin{array}
[c]{cc}%
A_{\eta} & B_{\eta}\\
0_{m\times\left(  n-m\right)  } & -\Lambda
\end{array}
\right]  \left[
\begin{array}
[c]{c}%
\eta\\
y
\end{array}
\right]  +\left[
\begin{array}
[c]{c}%
0_{\left(  n-m\right)  \times\left(  n-m\right)  }\\
C^{T}B
\end{array}
\right]  \left(  h\left(  t,u\right)  -K^{T}x+\sigma\left(  t,x\right)
\right)  \label{perturbedsystem_tran1}%
\end{equation}
where $\eta\in%
%TCIMACRO{\U{211d} }%
%BeginExpansion
\mathbb{R}
%EndExpansion
^{n-m}$ is an internal part and $y\in%
%TCIMACRO{\U{211d} }%
%BeginExpansion
\mathbb{R}
%EndExpansion
^{m}$ is an external part. Since $A$ is stable, $A_{\eta}\in%
%TCIMACRO{\U{211d} }%
%BeginExpansion
\mathbb{R}
%EndExpansion
^{\left(  n-m\right)  \times\left(  n-m\right)  }$ is stable too. In fact, a
minimum-phase MIMO system is designed by the new output matrix.

\subsection{Additive State Decomposition}

Consider the system (\ref{externalsystem}) as the original system. The primary
system is chosen as follows:%
\begin{equation}
\dot{y}_{p}=-\Lambda y_{p}+C^{T}Bu,y_{p}\left(  0\right)  =0.
\label{perturbedsystem_tran_pri}%
\end{equation}
Then the secondary system is determined by the original system
(\ref{perturbedsystem_tran}) and the primary system
(\ref{perturbedsystem_tran_pri}) with the rule (\ref{Dif_Sec_Sys}), resulting
in%
\begin{equation}
\dot{y}_{s}=-\Lambda y_{s}+C^{T}B\left(  -u+h\left(  t,u\right)  -K^{T}\left(
{x}\right)  +\sigma\left(  t,{x}\right)  \right)  ,y_{s}\left(  0\right)
=C^{T}x_{0}. \label{perturbedsystem_tran_scd}%
\end{equation}
According to (\ref{Gen_RelationDif}), it follows%
\begin{equation}
y=y_{p}+y_{s}. \label{relationship}%
\end{equation}
Define a transfer function $G\left(  s\right)  =\left(  sI_{m}+\Lambda\right)
^{-1}C^{T}B.$ Then, rearranging (\ref{perturbedsystem_tran_pri}%
)-(\ref{relationship}) results in%
\begin{align}
\dot{y}_{p}  &  =-\Lambda y_{p}+C^{T}Bu,y_{p}\left(  0\right)  =0.\nonumber\\
y  &  =y_{p}+d_{l} \label{modifiedperturbedsystem}%
\end{align}
where $d_{l}$ $=$ $C^{T}x_{s}$ $=G\left(  -u+h\left(  t,u\right)
-K^{T}x+\sigma\left(  t,x\right)  \right)  +e^{-\Lambda t}C^{T}x_{0}\ $is
called the lumped disturbance. Furthermore, (\ref{modifiedperturbedsystem}) is
written as%
\begin{equation}
y=Gu+d_{l}. \label{modifiedperturbedsystem1}%
\end{equation}
The lumped disturbance $d_{l}$ includes\textbf{\ }uncertainties, disturbance
and input. Fortunately, since $y_{p}=Gu$ and the output $y$ are known, the
lumped disturbance $d_{l}$ can be observed exactly by%
\begin{equation}
\hat{d}_{l}=y-Gu. \label{dis}%
\end{equation}
It is easy to see that $\hat{d}_{l}\equiv d_{l}.$

\subsection{Dynamic Inversion Control}

So far, by ASD, the uncertain system (\ref{perturbedsystem})\ has been
transformed into an uncertainty-free system (\ref{modifiedperturbedsystem1})
but subject to a lumped disturbance, which is shown in Fig.1.
\begin{figure}[h]
\begin{center}
\includegraphics[
scale=0.85 ]{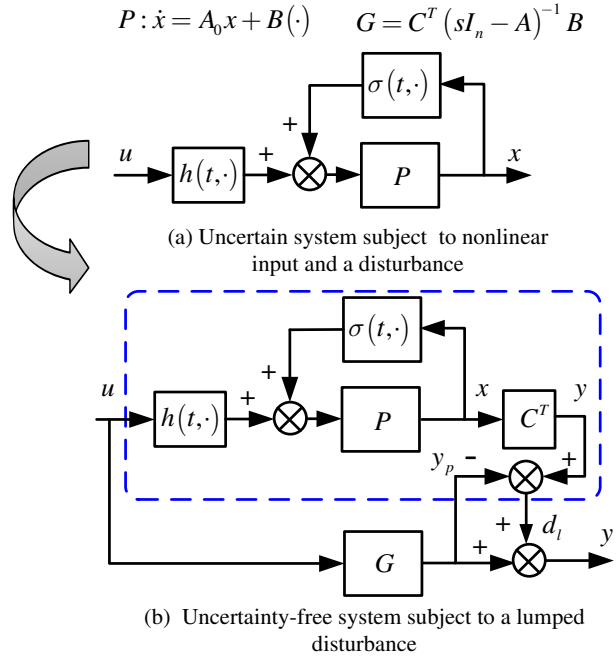}
\end{center}
\caption{Model Transformation}%
\end{figure}

For the system (\ref{modifiedperturbedsystem1}), since $G$ is minimum-phase
and known, the dynamic inversion tracking controller design is represented as
follows:%
\begin{equation}
u=-G^{-1}\hat{d}_{l}. \label{simcontroller}%
\end{equation}
Substituting (\ref{simcontroller}) into (\ref{modifiedperturbedsystem1})
results in%
\begin{align*}
y  &  =-GG^{-1}\hat{d}_{l}+d_{l}\\
&  =-\hat{d}_{l}+d_{l}=0
\end{align*}
where $\hat{d}\equiv d_{l}$ is utilized. As a result, perfect tracking is
achieved. However, the proposed controller (\ref{simcontroller}) cannot be
realized. By introducing a low-pass filter matrix $Q$, the controller
(\ref{simcontroller}) is modified as follows:%
\begin{equation}
u=-QG^{-1}\hat{d}_{l} \label{modifiedcontroller}%
\end{equation}
which has the simple structure shown in Fig.2. Furthermore, substituting
(\ref{dis}) into (\ref{modifiedcontroller}) results in
\[
u=-QG^{-1}\left(  y-Gu\right)  .
\]
Then, it can be further written as%
\begin{equation}
u=-\left(  I_{m}-Q\right)  ^{-1}QG^{-1}C^{T}x \label{closedform}%
\end{equation}
\begin{figure}[h]
\begin{center}
\includegraphics[
scale=0.85 ]{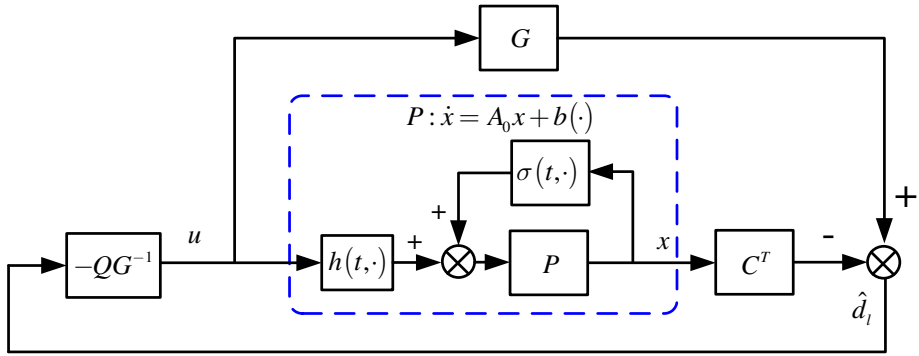}
\end{center}
\caption{Structure of the ASD Dynamic Inversion Stabilized Controller}%
\end{figure}

If det$\left(  C^{T}B\right)  \neq0$, then the controller above is realizable.
By employing the controller (\ref{modifiedcontroller}), the output becomes
\begin{equation}
y=\left(  I_{m}-Q\right)  \hat{d}_{l}. \label{trackingerror}%
\end{equation}
Since $Q$ is a low-pass filter matrix, and the low-frequency range is often
dominant in a signal, it is expected that the output will be attenuated by the
transfer function $I_{m}-Q$. A detailed analysis is given in the following section.

\textbf{Remark 3}. By the proposed output matrix redefinition, the considered
MIMO uncertainty system is transformed into a MIMO minimum-phase system with
relative degree one. By ASD, it is further transformed into a simple transfer
function, namely (\ref{modifiedperturbedsystem1}). Owing to the simple
transfer function, the controller design for (\ref{modifiedperturbedsystem1})
is straightforward by the idea of dynamic inversion.

\subsection{Performance Analysis}

Since the\ lumped disturbance $d_{l}$ involves\textbf{\ }the\textbf{\ }input
$u$, the resultant closed-loop system may be unstable. Next, some conditions
are given to guarantee that the control input $u$ is bounded. Substituting
$\hat{d}_{l}$ into (\ref{modifiedcontroller}) results in%
\begin{align}
u  &  =-QG^{-1}\left[  G\left(  -u+h\left(  t,u\right)  -K^{T}x+\sigma\left(
t,x\right)  \right)  +e^{-\Lambda t}C^{T}x_{0}\right] \nonumber\\
&  =Q\left(  u-h\left(  t,u\right)  +K^{T}x-\sigma\left(  t,x\right)  \right)
-QG^{-1}e^{-\Lambda t}C^{T}x_{0}. \label{modifiedcontroller1}%
\end{align}
Multiplying $Q^{-1}$ on both sides of (\ref{modifiedcontroller1}) yields%
\begin{equation}
Q^{-1}u=u-h\left(  t,u\right)  +K^{T}x-\sigma\left(  t,x\right)
+\xi\label{modifiedcontroller2}%
\end{equation}
where $\xi=-G^{-1}e^{-\Lambda t}C^{T}x_{0}$. Since $G=\left(  sI_{m}%
+\Lambda\right)  ^{-1}C^{T}B$, the term $\xi$ will tend to zero exponentially.
A simple way is to choose $Q=\frac{1}{\epsilon s+1},$ where $\epsilon>0$ can
be considered as a singular perturbation parameter. By the filter $Q$,
(\ref{modifiedcontroller2}) is further written as%
\begin{equation}
\epsilon\dot{u}=-h\left(  t,u\right)  +K^{T}x-\sigma\left(  t,x\right)  +\xi.
\label{modifiedcontroller3}%
\end{equation}
The following theorem will give an explicit bound on $\epsilon$, below which
the stability of closed-loop dynamics forming by (\ref{perturbedsystem_tran})
and (\ref{modifiedcontroller3}) can be guaranteed.

\textbf{Theorem 2}. Suppose i) \textit{Assumptions 1-3 }hold, ii)\textit{\ }%
the controller is designed as (\ref{modifiedcontroller}) with $Q=\frac
{1}{\epsilon s+1},$ iii) $\epsilon$ satisfies%
\begin{equation}
0<\epsilon<\frac{\underline{l}_{h_{u}}}{\gamma_{1}+\frac{2}{\gamma_{0}}\left(
\gamma_{2}+l_{\sigma_{t}}\right)  ^{2}} \label{condition}%
\end{equation}
where%
\begin{align}
\gamma_{0}  &  =\lambda_{\min}\left(  M\right) \nonumber\\
\gamma_{1}  &  =2\left(  \left\Vert K\right\Vert +l_{\sigma_{x}}\right)
\left\Vert b\right\Vert +2\frac{l_{h_{t}}}{\underline{l}_{h_{u}}}\nonumber\\
\gamma_{2}  &  =\left\Vert P\right\Vert \left\Vert b\right\Vert +\left\Vert
A\right\Vert \left(  \left\Vert K\right\Vert +l_{\sigma_{x}}\right)
+\left\Vert K\right\Vert +k_{\sigma}. \label{parameters}%
\end{align}
Then the state of system (\ref{perturbedsystem}) is uniformly ultimately
bounded with respect to the bound $\sqrt{\frac{1}{\eta\left(  \epsilon\right)
}\frac{\epsilon}{\underline{l}_{h_{u}}}}\left(  \frac{l_{h_{t}}}{\underline
{l}_{h_{u}}}\delta_{\sigma}+d_{\sigma}\right)  ,$ where $\eta\left(
\epsilon\right)  =\min(\frac{\gamma_{0}}{2\lambda_{\max}\left(  P\right)
},\frac{1}{\epsilon}\underline{l}_{h_{u}}-\gamma_{1}-\frac{2}{\gamma_{0}%
}\left(  \gamma_{2}+l_{\sigma_{t}}\right)  ^{2}).$ Furthermore, if $l_{h_{t}%
}\delta_{\sigma}\left(  t\right)  \rightarrow0$ and $d_{\sigma}\left(
t\right)  \rightarrow0$, then the state $x\left(  t\right)  \rightarrow0$ as
$t\rightarrow\infty.$

\textit{Proof}. See Appendix.

\textbf{Remark 4}. According to \textit{Theorem 2},\textit{\ }if the
uncertainties are time invariant, namely $l_{h_{t}}=0$ and $d_{\sigma}=0,$
then the proposed controller results in asymptotic stability. Also, from
\textit{Theorem 2}, a sufficiently small $\epsilon$ will satisfy
(\ref{condition}) and the ultimate bound will be reduced by decreasing
$\epsilon.$ However, it should be pointed out that a small $\epsilon$ in turn
will result in a reduced stability of the closed-loop system, namely the roots
are closer to the imaginary axis. For example, consider a simple situation
where $h\left(  t,u\right)  =u\left(  t-\tau\right)  .$ Given $\tau>0$, the
dynamical system $\epsilon\dot{u}=-u\left(  t-\tau\right)  $ will lose
stability on choosing a sufficiently small $\epsilon$ no matter how small the
delay $\tau$ is (The characteristic equation of $\epsilon\dot{u}=-u\left(
t-\tau\right)  $ is $\epsilon s+e^{-s\tau}=0,$ which can be approximated by
$\left(  \epsilon-\tau\right)  s+1=0.$ Therefore, one solution of the
characteristic equation is $s\approx\frac{1}{\tau-\epsilon}.$ If
$\epsilon<\tau$, then $s>0$, namely the dynamical system $\epsilon\dot
{u}=-u\left(  t-\tau\right)  $ is unstable). Therefore, an appropriate
$\epsilon>0\ $should be chosen to achieve a tradeoff between tracking
performance and robustness. The parameters $l_{h_{t}},\underline{l}_{h_{u}%
},\overline{l}_{h_{u}},k_{\sigma},\delta_{\sigma},l_{\sigma_{x}},l_{\sigma
_{t}},d_{\sigma}$ need not be known, and $\epsilon$ is chosen as large as
possible consistent with the state subject to an acceptable uniform ultimate
bound according to practical requirements. From the above analysis, the design
procedure is summarized as follows.%
\[%
\begin{tabular}
[c]{|c||l|}\hline
\multicolumn{2}{|c|}{\textbf{Procedure}}\\\hline
\textit{Step 1} & \textit{Design a state feedback gain }$K\in%
%TCIMACRO{\U{211d} }%
%BeginExpansion
\mathbb{R}
%EndExpansion
^{n\times m}$\textit{ such that }$A=A_{0}+BK^{T}$\textit{ is stable with}\\
& $n$\textit{ negative real eigenvalues denoted by }$-\lambda_{i}<0,$\textit{
}$i=1,\cdots,n.$\\\hline
\textit{Step 2} & \textit{Define }$C=\left[
\begin{array}
[c]{ccc}%
c_{1} & \cdots & c_{m}%
\end{array}
\right]  \in%
%TCIMACRO{\U{211d} }%
%BeginExpansion
\mathbb{R}
%EndExpansion
^{n\times m},$\textit{ where }$c_{i}\in%
%TCIMACRO{\U{211d} }%
%BeginExpansion
\mathbb{R}
%EndExpansion
^{n}$ \textit{are independent unit}\\
& \textit{eigenvectors corresponding to eigenvalues }$-\lambda_{i}$\textit{ of
}$A^{T},$\textit{ }$i=1,\cdots,m,$\textit{ respectively.}\\\hline
\textit{Step 3} & \textit{Design (\ref{modifiedcontroller}) or
(\ref{closedform}) with }$G=\left(  sI_{m}+\Lambda\right)  ^{-1}C^{T}%
B$\textit{ and }$Q=\frac{1}{\epsilon s+1}$\textit{.}\\\hline
\textit{Step 4.} & \textit{Choose the appropriate }$\epsilon>0\ $\textit{in
practice (see text).}\\\hline
\end{tabular}
\ \ \ \
\]

\section{Numerical Simulations}

To demonstrate its effectiveness, the proposed control method is applied to
two existing problems in \cite{Young(2007)},\cite{Hsu(1998)}\ for comparison
by numerical simulations.

\subsection{An Uncertain SISO System}

As in \cite{Hsu(1998)}, the following uncertain dynamics are considered
\begin{equation}
\dot{x}\left(  t\right)  =A_{0}x\left(  t\right)  +B\left(  \Phi\left(
u\left(  t\right)  \right)  +e\left(  x,t\right)  \right)  \label{Nonlinear}%
\end{equation}
where%
\begin{align*}
A_{0}  &  =\left[
\begin{array}
[c]{ccc}%
0 & 1 & 0\\
0 & 0 & 1\\
-1 & -3 & -1
\end{array}
\right]  ,B=\left[
\begin{array}
[c]{c}%
0\\
0\\
1
\end{array}
\right] \\
\Phi\left(  u\left(  t\right)  \right)   &  =\left(  0.5+0.3\sin u\left(
t\right)  +e^{0.2\left\vert \cos u\left(  t\right)  \right\vert }\right)
u\left(  t\right) \\
e\left(  x,t\right)   &  =\left(  0.3+0.2\cos x_{1}\right)  \sqrt{x_{1}%
^{2}+x_{2}^{2}+x_{3}^{2}}-0.5\sin x_{2}%
\end{align*}
The objective is to drive $x\left(  t\right)  \rightarrow0$ as $t\rightarrow
\infty$.

For the dynamics (\ref{Nonlinear}), according to \textit{Procedure}, the
following design is given.

\textit{Step 1}. According to the procedure, design $K=\left[
\begin{array}
[c]{ccc}%
-5 & -8 & -5
\end{array}
\right]  ^{T}$ resulting in $A=A_{0}+BK^{T}$ with $3$ different negative real
eigenvalues $-1,-2,-3$.

\textit{Step 2}. Selecting eigenvectors of $A^{T}$ corresponding to its
eigenvalues $-1\ $results in $C=\left[
\begin{array}
[c]{ccc}%
6 & 5 & 1
\end{array}
\right]  ^{T}.$

\textit{Step 3}. Design controller $u=-QG^{-1}\hat{d}_{l}=-\frac{1}{\epsilon
}\left(  1+\frac{1}{s}\right)  C^{T}x,$ where $G=\frac{1}{s+1}$,$\ Q=\frac
{1}{\epsilon s+1},$ $\hat{d}_{l}=C^{T}x-Gu$.

\textit{Step 4}. \textit{Choose }$\epsilon=0.1.$

The range of the control input $u$ is chosen as $\left[  -5,5\right]  \ $in
practice. Driven by the designed controller, the control performance is shown
in Fig. 3. As shown, all states converge to zero. Moreover, the control input
is continuous and bounded. The control performance by the sliding mode
controller proposed in \cite{Hsu(1998)} is shown in Fig. 4. The index
$E\left(  t\right)  =%
%TCIMACRO{\dint \nolimits_{0}^{t}}%
%BeginExpansion
{\displaystyle\int\nolimits_{0}^{t}}
%EndExpansion
\left\vert \dot{u}\left(  s\right)  \right\vert ds$ is introduced to represent
the energy cost. It is easy to observe that our proposed controller saves more
energy compared with the sliding mode controller proposed in \cite{Hsu(1998)}%
.\begin{figure}[h]
\begin{center}
\includegraphics[
scale=0.65 ]{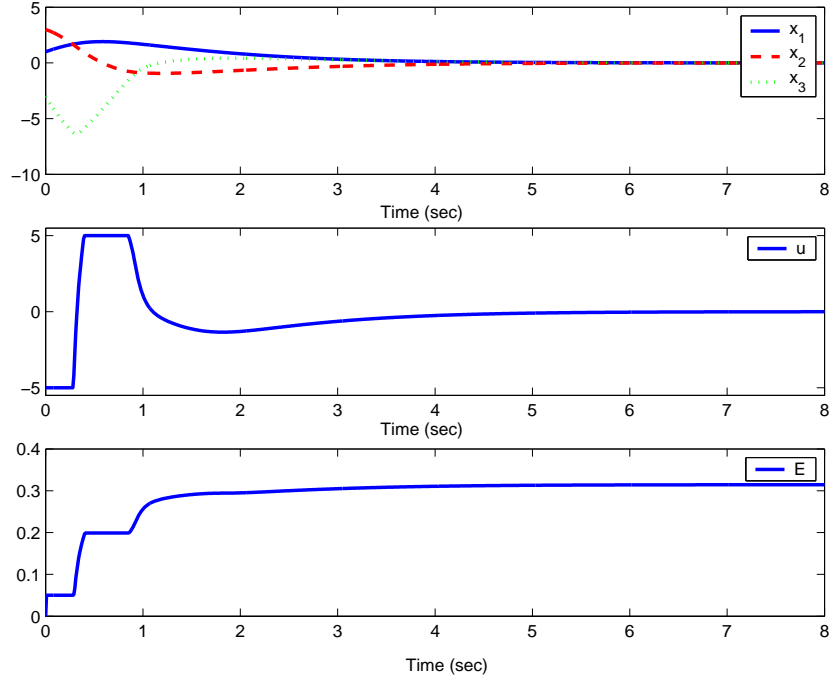}
\end{center}
\caption{Stabilization Performance of the SISO Dynamics Driven by the ASD
Dynamic Inversion Stabilized Controller}%
\end{figure}\begin{figure}[h]
\begin{center}
\includegraphics[
scale=0.65 ]{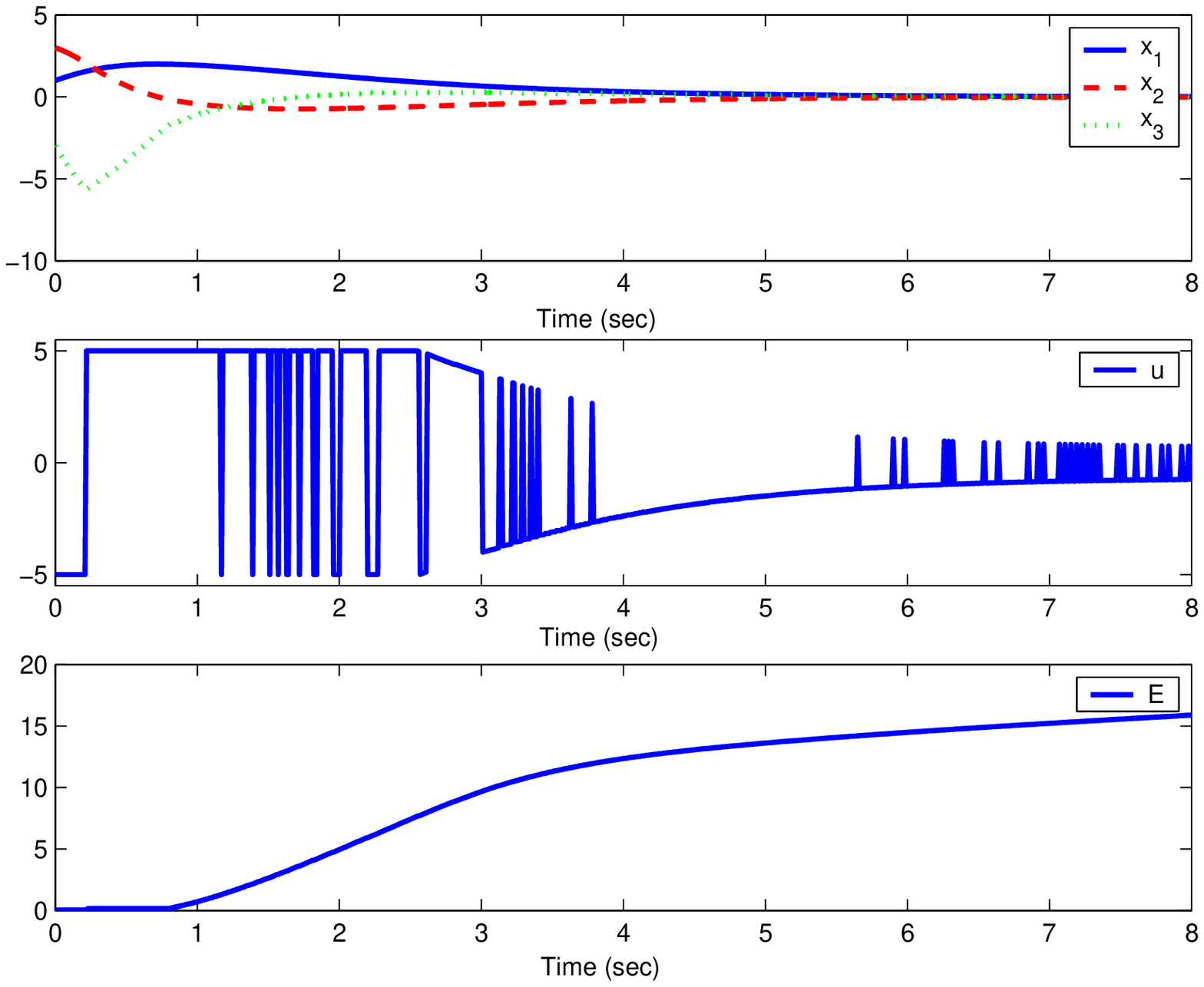}
\end{center}
\caption{Stabilization Performance of the SISO Dynamics Driven by the Sliding
Mode Controller Proposed in [10]}%
\end{figure}

\subsection{An Uncertain MIMO System}

As in \cite{Young(2007)}, the lateral/directional baseline model of an F-16
from \cite{Stevens(1992)} flying at sea level with an airspeed of $502$ $ft/s$
and angle of attack of $2.11$ deg is used. Denote the angle of sideslip, the
roll angle, the stability axis roll and yaw rates, aileron and rudder control
by $\beta,\phi,p_{s},r_{s},\delta_{a},\delta_{r}$ respectively. The full
roll/yaw dynamics in state space form gives%
\begin{equation}
\underset{\dot{x}\left(  t\right)  }{\underbrace{\left[
\begin{array}
[c]{c}%
\dot{\beta}\left(  t\right) \\
\dot{\phi}\left(  t\right) \\
\dot{p}_{s}\left(  t\right) \\
\dot{r}_{s}\left(  t\right)
\end{array}
\right]  }}=A_{0}\underset{x\left(  t\right)  }{\underbrace{\left[
\begin{array}
[c]{c}%
\beta\left(  t\right) \\
\phi\left(  t\right) \\
p_{s}\left(  t\right) \\
r_{s}\left(  t\right)
\end{array}
\right]  }}+B\left[
\begin{array}
[c]{c}%
\delta_{a}\left(  t\right)  +f_{1}\left(  \beta\left(  t\right)  ,p_{s}\left(
t\right)  ,r_{s}\left(  t\right)  ,\delta_{a}\left(  t\right)  \right) \\
\delta_{r}\left(  t\right)  +f_{2}\left(  \beta\left(  t\right)  ,p_{s}\left(
t\right)  ,r_{s}\left(  t\right)  ,\delta_{r}\left(  t\right)  \right)
\end{array}
\right]  . \label{roll/yaw}%
\end{equation}
Here
\begin{align*}
A_{0}  &  =\left[
\begin{array}
[c]{cccc}%
-0.3220 & 0.064 & 0.0364 & -0.9917\\
0 & 0 & 1 & 0.0393\\
-30.6490 & 0 & -3.6784 & 0.6646\\
8.5395 & 0 & -0.0254 & -0.4764
\end{array}
\right]  ,B=\left[
\begin{array}
[c]{cc}%
0 & 0\\
0 & 0\\
-0.7331 & 0.1315\\
-0.0319 & -0.0620
\end{array}
\right] \\
f_{1}  &  =\left(  \left(  1-C_{1}\right)  e^{-\frac{\left(  \beta-\beta
_{0}\right)  ^{2}}{2\sigma_{1}^{2}}}+C_{1}\right)  \left(  \tanh\left(
\delta_{a}+h_{1}\right)  +\tanh\left(  \delta_{a}-h_{1}\right)  +0.001\delta
_{a}\right) \\
&  +D_{1}\cos\left(  A_{1}p_{s}-\omega_{1}\right)  \sin\left(  A_{2}%
r_{s}-\omega_{2}\right)  +D_{2}\\
f_{2}  &  =\left(  \left(  1-C_{2}\right)  e^{-\frac{\left(  \beta-\beta
_{0}\right)  ^{2}}{2\sigma_{2}^{2}}}+C_{2}\right)  \left(  \tanh\left(
\delta_{r}+h_{2}\right)  +\tanh\left(  \delta_{a}-h_{2}\right)  +0.001\delta
_{r}\right) \\
&  +D_{3}\cos\left(  A_{3}p_{s}-\omega_{3}\right)  \sin\left(  A_{4}%
r_{s}-\omega_{4}\right)  +D_{4}.
\end{align*}
where $A_{1}=0.33,$ $A_{2}=0.195,$ $A_{3}=0.45,$ $A_{4}=1.85,$ $D_{1}=0.295,$
$D_{2}=-0.0865,$ $D_{3}=0.055,$ $D_{4}=-0.007,$ $w_{1}=1.6,$ $w_{2}=0,$
$w_{3}=-1.9,$ $w_{4}=0,$ $C_{1}=0.3,$ $C_{2}=0.3,$ $h_{1}=7,$ $h_{2}=2.7,$
$\delta_{1}=0.25,$ $\delta_{2}=0.25,$ $\beta_{0}=0.$ Readers are refered to
\cite{Young(2007)} for details. The dynamics (\ref{roll/yaw}) can be
formulated as (\ref{perturbedsystem}). The objective is to drive $x\left(
t\right)  \rightarrow0$ as $t\rightarrow\infty$.

For the dynamics (\ref{roll/yaw}), according to the design procedure at the
end of Section III.C, the following design is given.

\textit{Step 1}. According to the procedure, design
\[
K=\left[
\begin{array}
[c]{cc}%
-27.5037 & 93.4020\\
14.2953 & 35.0244\\
4.5010 & 13.9005\\
12.7039 & 58.8096
\end{array}
\right]
\]
resulting in $A=A_{0}+BK^{T}$ with the $4$ negative real eigenvalues
$-1,-2,-3,-4$.

\textit{Step 2}. Selecting eigenvectors of $A^{T}$ corresponding to its
eigenvalues $-1,-2\ $results in%
\[
C=\left[
\begin{array}
[c]{cc}%
0.6537 & -0.5473\\
0.6819 & 0.7985\\
0.2285 & 0.1961\\
-0.2354 & 0.1561
\end{array}
\right]  .
\]

\textit{Step 3}. Design controller
\begin{equation}
u=\left[
\begin{array}
[c]{c}%
\delta_{a}\\
\delta_{r}%
\end{array}
\right]  =-QG^{-1}\hat{d}_{l}=-\left(  C^{T}B\right)  ^{-1}\left[
\begin{array}
[c]{cc}%
\frac{s+1}{\epsilon s} & 0\\
0 & \frac{s+2}{\epsilon s}%
\end{array}
\right]  C^{T}x \label{con_aircraft}%
\end{equation}
where $G=\left[
\begin{array}
[c]{cc}%
\frac{1}{s+1} & 0\\
0 & \frac{1}{s+2}%
\end{array}
\right]  C^{T}B$,$\ Q=\frac{1}{\epsilon s+1},$ $\hat{d}_{l}=C^{T}x-Gu$.

\textit{Step 4}. Choose the appropriate\textit{ }$\epsilon=0.2.$

The range of the control input $\delta_{a},\delta_{r}$ is chosen as $\left[
-20\deg,20\deg\right]  \ $in practice. Driven by (\ref{con_aircraft}), the
control performance is shown in Fig. 5. As shown, all states converge to zero.
Moreover, the control input is continuous and bounded. Here the information of
nonlinear terms $f_{1}$ and $f_{2}$ is not required, let alone learn the
parameters. So, compared with controller proposed in \cite{Young(2007)}, the
controller design and controller structure are both simpler.\begin{figure}[h]
\begin{center}
\includegraphics[
scale=0.65 ]{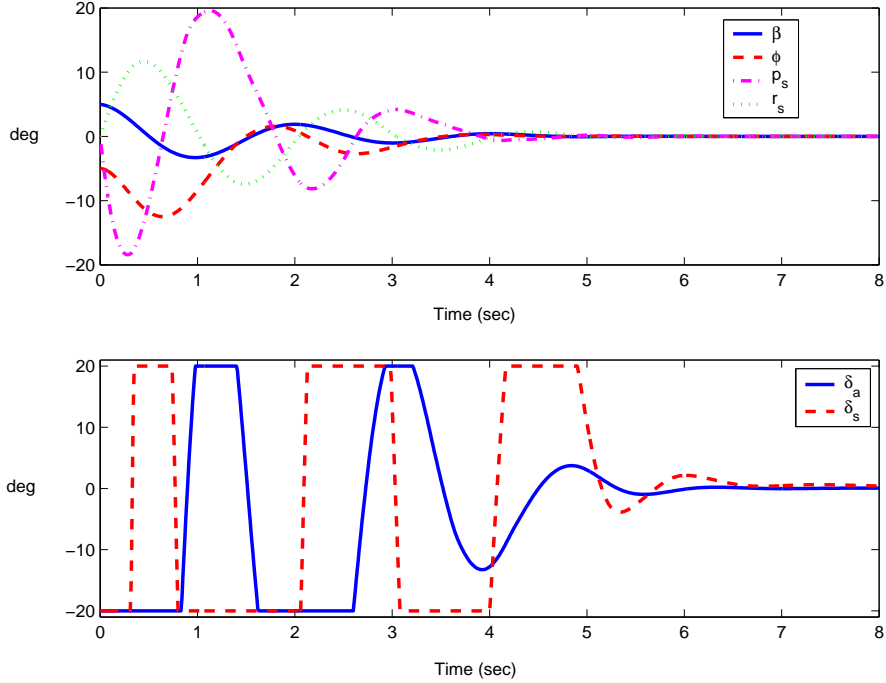}
\end{center}
\caption{Stabilization Performance of F-16 Roll/Yaw Dynamics Driven by the ASD
Dynamic Inversion Stabilized Controller}%
\end{figure}

\section{An Application: Attitude Control of A Quadrotor}

In this section, in order to show its practicability, the proposed ASD dynamic
inversion stabilized controller is applied to attitude control of a quadrotor
when its inertia moment matrix is subject to a large uncertainty.

\subsection{Problem Formulation}

By taking actuator dynamics into account, the linear roll model of the
quadrotor around hover conditions is%
\begin{equation}
\underset{\dot{x}_{\phi}}{\underbrace{\left[
\begin{array}
[c]{c}%
\dot{\phi}\left(  t\right) \\
\dot{p}\left(  t\right) \\
\dot{L}\left(  t\right)
\end{array}
\right]  }}=\underset{A_{0}}{\underbrace{\left[
\begin{array}
[c]{ccc}%
0 & 1 & 0\\
0 & 0 & 1\\
0 & 0 & -\omega
\end{array}
\right]  }}\underset{x_{\phi}}{\underbrace{\left[
\begin{array}
[c]{c}%
\phi\left(  t\right) \\
p\left(  t\right) \\
L\left(  t\right)
\end{array}
\right]  }}+\underset{B_{0}}{\underbrace{\left[
\begin{array}
[c]{c}%
0\\
0\\
\omega
\end{array}
\right]  }}J_{\phi}^{-1}\tau_{\phi} \label{roll channel}%
\end{equation}
where $\omega$ is the actuator bandwidth, $\tau_{\phi}$ is the command torque
to be selected, and $x_{\phi}=\left[
\begin{array}
[c]{ccc}%
\phi & p & L
\end{array}
\right]  ^{T}\in%
%TCIMACRO{\U{211d} }%
%BeginExpansion
\mathbb{R}
%EndExpansion
^{3}\ $with $\phi,p,L\in%
%TCIMACRO{\U{211d} }%
%BeginExpansion
\mathbb{R}
%EndExpansion
$ being the angle, angular velocity and torque of the roll channel in the
body-fixed frame, respectively. In practice, $\phi,p$ can be measured but $L$
cannot be. According to this, $L$ is estimated by $\dot{\hat{L}}=-\omega
\hat{L}+\omega\tau_{\phi},\hat{L}\left(  0\right)  =0,$ taken as the true
measurement for simplicity. In this sense, $x_{\phi}$ is known. Let $J\in%
%TCIMACRO{\U{211d} }%
%BeginExpansion
\mathbb{R}
%EndExpansion
^{3\times3}$ be the inertia moment matrix of the quadrotor, $x_{\theta
}=\left[
\begin{array}
[c]{ccc}%
\theta & q & M
\end{array}
\right]  ^{T}\in%
%TCIMACRO{\U{211d} }%
%BeginExpansion
\mathbb{R}
%EndExpansion
^{3}$ and$\ x_{\psi}=\left[
\begin{array}
[c]{ccc}%
\theta & q & N
\end{array}
\right]  ^{T}\in%
%TCIMACRO{\U{211d} }%
%BeginExpansion
\mathbb{R}
%EndExpansion
^{3}.$ Here $\theta,\psi\in%
%TCIMACRO{\U{211d} }%
%BeginExpansion
\mathbb{R}
%EndExpansion
$ are pitch and yaw angle respectively, while $q,r$\ are respectively their
angular velocity in the body-fixed frame. The torques $M,N\in%
%TCIMACRO{\U{211d} }%
%BeginExpansion
\mathbb{R}
%EndExpansion
\ $represent the airframe pitch and yaw torque, respectively. Similar to
(\ref{roll channel}), the linear attitude model of the quadrotor is expressed
as%
\begin{equation}
\dot{x}=\bar{A}_{0}x+BJ^{-1}\tau,x\left(  0\right)  =x_{0}. \label{hexacopter}%
\end{equation}
Here $x=\left[
\begin{array}
[c]{ccc}%
x_{\phi} & x_{\theta} & x_{\psi}%
\end{array}
\right]  ^{T}\in%
%TCIMACRO{\U{211d} }%
%BeginExpansion
\mathbb{R}
%EndExpansion
^{9}$ is the state and $\tau=\left[
\begin{array}
[c]{ccc}%
\tau_{\phi} & \tau_{\theta} & \tau_{\psi}%
\end{array}
\right]  ^{T}\in%
%TCIMACRO{\U{211d} }%
%BeginExpansion
\mathbb{R}
%EndExpansion
^{3}.$ The system matrix and input matrix in (\ref{hexacopter}) are%
\begin{equation}
\bar{A}_{0}=\left[
\begin{array}
[c]{ccc}%
A_{0} & 0_{3\times3} & 0_{3\times3}\\
0_{3\times3} & A_{0} & 0_{3\times3}\\
0_{3\times3} & 0_{3\times3} & A_{0}%
\end{array}
\right]  \in%
%TCIMACRO{\U{211d} }%
%BeginExpansion
\mathbb{R}
%EndExpansion
^{9\times9},B=\left[
\begin{array}
[c]{ccc}%
B_{0} & 0_{3\times1} & 0_{3\times1}\\
0_{3\times1} & B_{0} & 0_{3\times1}\\
0_{3\times1} & 0_{3\times1} & B_{0}%
\end{array}
\right]  \in%
%TCIMACRO{\U{211d} }%
%BeginExpansion
\mathbb{R}
%EndExpansion
^{9\times3}. \label{hexa_AB}%
\end{equation}

In practice, the inertia moment matrix $J\in%
%TCIMACRO{\U{211d} }%
%BeginExpansion
\mathbb{R}
%EndExpansion
^{3\times3}$, related to the position and weight of payload such as
instruments and batteries, is often difficult to determine. Moreover, the
payload of the quadrotor is often time-varying owing to fuel consumption or
pesticide spraying. Therefore, compared with the true inertia moment matrix,
the real inertia moment matrix $J$ may have a large uncertainty. Assume the
nominal inertia moment matrix to be $J_{0}\in%
%TCIMACRO{\U{211d} }%
%BeginExpansion
\mathbb{R}
%EndExpansion
^{3\times3}.$ By employing it, a controller is designed to stabilize the
attitude (\ref{hexacopter}) as%
\begin{equation}
\tau=J_{0}\left(  \bar{K}^{T}x+u\right)  \label{torque}%
\end{equation}
where $\bar{K}\in%
%TCIMACRO{\U{211d} }%
%BeginExpansion
\mathbb{R}
%EndExpansion
^{9\times3}$ and $u$ will be specified later. Then (\ref{hexacopter}) can be
cast in the form of (\ref{perturbedsystem}) as%
\begin{equation}
\dot{x}=A_{0}x+B\left(  h\left(  t,u\right)  +\sigma\left(  t,x\right)
\right)  ,x\left(  0\right)  =x_{0} \label{control torque}%
\end{equation}
where $A_{0}=\bar{A}_{0}+B\bar{K}^{T},$ $h\left(  t,u\right)  =J^{-1}J_{0}u,$
and $\sigma\left(  t,x\right)  =\left(  J^{-1}J_{0}-I_{3}\right)  \bar{K}%
^{T}x.$ To accord with the \textit{Step 1} of the proposed control procedure,
choose%
\[
\bar{K}=\left[
\begin{array}
[c]{ccc}%
K_{0} & 0_{3\times1} & 0_{3\times1}\\
0_{3\times1} & K_{0} & 0_{3\times1}\\
0_{3\times1} & 0_{3\times1} & K_{0}%
\end{array}
\right]  ,K_{0}=\left[
\begin{array}
[c]{c}%
-3.0\\
-4.2\\
-0.27
\end{array}
\right]  .
\]
The resultant $A_{0}=\bar{A}_{0}+B\bar{K}^{T}$ is stable with $\det\left(
sI_{9}-A_{0}\right)  =\left(  s+15\right)  ^{3}\left(  s+3\right)  ^{3}\left(
s+1\right)  ^{3}.$

\subsection{ASD Dynamic Inversion Control}

For system (\ref{hexacopter}), according to the procedure in Section III, the
following design steps are given.

\textit{Step 1.} Thanks to the designed $A_{0}$ above, \textit{Step 1} is
skipped. As a result, the matrix $A=A_{0}$ is stable and $\det\left(
sI_{9}-A\right)  =\left(  s+15\right)  ^{3}\left(  s+3\right)  ^{3}\left(
s+1\right)  ^{3}$.

\textit{Step 2}. The output matrix $C$ is redefined by%
\begin{equation}
C=\left[
\begin{array}
[c]{ccc}%
C_{0} & 0_{3\times1} & 0_{3\times1}\\
0_{3\times1} & C_{0} & 0_{3\times1}\\
0_{3\times1} & 0_{3\times1} & C_{0}%
\end{array}
\right]  ,C_{0}=\left[
\begin{array}
[c]{c}%
0.9283\\
0.3713\\
0.0206
\end{array}
\right]  \label{hexa_C}%
\end{equation}
with $C^{T}A=-C^{T}$.

\textit{Step 3. }Design%
\[
u=-\left(  1-Q\right)  ^{-1}QG^{-1}C^{T}x
\]
where $G=\frac{1}{s+1}C^{T}B$ and $Q=\frac{1}{\epsilon s+1}$. Rearranging the
control term above yields%
\begin{equation}
u=-\frac{1}{\epsilon}\left(  C^{T}B\right)  ^{-1}C^{T}x-\frac{1}{\epsilon
s}\left(  C^{T}B\right)  ^{-1}C^{T}x. \label{u}%
\end{equation}

\textit{Step 4.} Choose the appropriate $\epsilon$.

\subsection{Experiment}

The experiment is performed on a Quanser Qball-X4, a quadrotor developed by
the Quanser company. Its nominal inertia matrix is $J_{0}=$diag$\left(
0.03,0.03,0.04\right)  $ kg$\cdot$m$^{2}$ and the actuator bandwidth is
$\omega=15$ rad/s.$\ $The parameter is chosen as $\epsilon=0.2$ for the
control term (\ref{u}). The proposed controller (\ref{torque}) is used only
for attitude control, while an existing position controller offered by the
Quanser company is retained. By using them, a hover control is expected to
perform for the Quanser Qball-X4. The experimental results are shown in Fig.
6. Then, in order to demonstrate the effectiveness of the proposed method, a
$0.145$ kg payload is attached to the $1.4$ kg Quanser Qball-X4 to change its
inertia moment matrix, shown in Fig. 7. With the same controller, the
experimental results are shown in Fig. 6. As shown, the proposed controller is
robust against the uncertainty in the inertia moment matrix.\begin{figure}[h]
\begin{center}
\includegraphics[
scale=0.65 ]{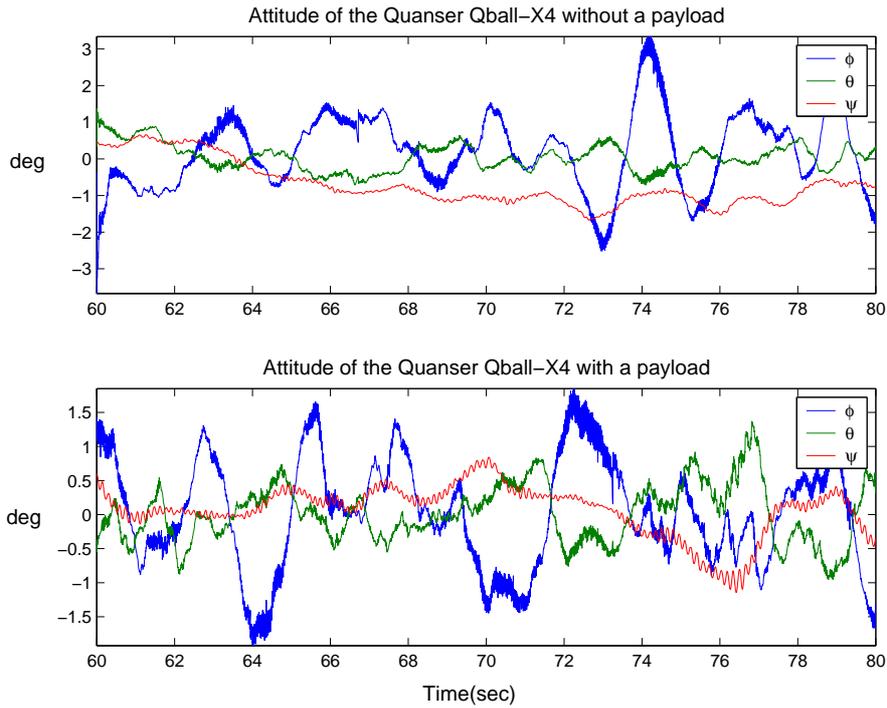}
\end{center}
\caption{Attitude Stabilization Performance of the ASD Dynamic Inversion
Stabilized Controller}%
\end{figure}\begin{figure}[h]
\begin{center}
\includegraphics[
scale=1 ]{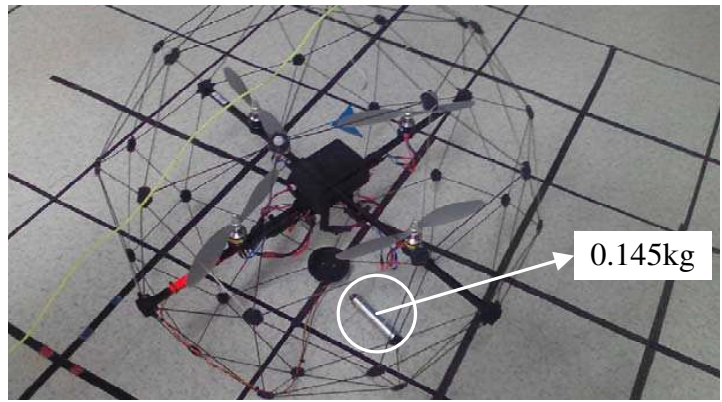}
\end{center}
\caption{The Quanser Qball-X4 is Attached a 0.145 kg Payload}%
\end{figure}

\section{Conclusion}

Stabilization for a class of MIMO systems is considered subject to
nonparametric time-varying uncertainties with respect to both state and input.
This study has three contributions: (i) an ASD dynamic inversion stabilized
control, which can solve the stabilization problem for a class of uncertain
MIMO systems, (ii) the definition of a new output matrix, which transforms
uncertain systems into minimum-phase systems with relative degree one, (iii) a
new ASD method, which further transforms the uncertain minimum-phase systems
into uncertainty-free systems with one observable lumped disturbance at the
output. From the simulations and the experiment, the proposed control scheme
has two salient features: less system information required and a simpler
design procedure with fewer tuning parameters.

\section*{Appendix: Proof of Theorem 2}

The following preliminary result is needed.

\textit{Lemma 1}\textbf{ }\cite{Dennis(1987)}. Let $F:$ $%
%TCIMACRO{\U{211d} }%
%BeginExpansion
\mathbb{R}
%EndExpansion
^{n}\rightarrow%
%TCIMACRO{\U{211d} }%
%BeginExpansion
\mathbb{R}
%EndExpansion
^{m}$ be continuously differentiable in an open convex set $D\subset%
%TCIMACRO{\U{211d} }%
%BeginExpansion
\mathbb{R}
%EndExpansion
^{n}$. For any $x,x+p\in D,$ $F\left(  x+p\right)  $ $-$ $F\left(  x\right)  $
$=$ $\int_{0}^{1}\left.  \frac{\partial F}{\partial z}\right\vert
_{z=x+sp}ds\cdot p.$

Denote $v=h\left(  t,u\right)  -K^{T}x+\sigma\left(  t,x\right)  .$ Then the
system (\ref{perturbedsystem_tran}) becomes%
\[
\dot{x}=Ax+Bv
\]
and the derivative of $\epsilon\dot{v}$ is calculated to be%
\begin{align*}
\epsilon\dot{v}  &  =\frac{\partial h}{\partial u}\epsilon\dot{u}%
+\epsilon\frac{\partial h}{\partial t}-\epsilon K^{T}\dot{x}+\frac
{\partial\sigma}{\partial x}\epsilon\dot{x}+\epsilon\frac{\partial\sigma
}{\partial t}\\
&  =\frac{\partial h}{\partial u}\left(  -v+\xi\right)  +\epsilon\left(
-K^{T}+\frac{\partial\sigma}{\partial x}\right)  \left(  Ax+Bv\right)
+\epsilon\frac{\partial h}{\partial t}+\epsilon\frac{\partial\sigma}{\partial
t}\text{\ ((\ref{modifiedcontroller3}) is used)}\\
&  =\left[  -\frac{\partial h}{\partial u}+\epsilon\left(  -K^{T}%
+\frac{\partial\sigma}{\partial x}\right)  B\right]  v+\epsilon\frac{\partial
h}{\partial t}+\epsilon\frac{\partial\sigma}{\partial t}+\frac{\partial
h}{\partial u}\xi+\epsilon\left(  -K^{T}+\frac{\partial\sigma}{\partial
x}\right)  Ax.
\end{align*}
Consequently, the closed-loop dynamics (\ref{modifiedcontroller3}) and
(\ref{perturbedsystem_tran}) are%
\begin{align}
\dot{x}  &  =Ax+bv\nonumber\\
\dot{v}  &  =\left[  -\frac{1}{\epsilon}\frac{\partial h}{\partial u}+\left(
-K^{T}+\frac{\partial\sigma}{\partial x}\right)  B\right]  v+\frac{\partial
h}{\partial t}+\frac{\partial\sigma}{\partial t}+\frac{1}{\epsilon}%
\frac{\partial h}{\partial u}\xi+\left(  -K^{T}+\frac{\partial\sigma}{\partial
x}\right)  Ax. \label{error dynamic}%
\end{align}
Choose a candidate Lyapunov function as follows:%
\[
V=x^{T}Px+v^{T}v
\]
where $0<P\in%
%TCIMACRO{\U{211d} }%
%BeginExpansion
\mathbb{R}
%EndExpansion
^{n\times n}\ $satisfies (\ref{PQ}). Taking the derivative of $V$ along the
solution of (\ref{error dynamic}) yields%
\begin{align*}
\dot{V}  &  =x^{T}\left(  PA+A^{T}P\right)  x+2x^{T}PBv+2v^{T}\left[
-\frac{1}{\epsilon}\frac{\partial h}{\partial u}+\left(  -K^{T}+\frac
{\partial\sigma}{\partial x}\right)  B\right]  v\\
&  \text{ \ \ }+2v^{T}\left(  -K^{T}+\frac{\partial\sigma}{\partial x}\right)
Ax+2v^{T}\left(  \frac{\partial h}{\partial t}+\frac{\partial\sigma}{\partial
t}+\frac{1}{\epsilon}\frac{\partial h}{\partial u}\xi\right) \\
&  =x^{T}\left(  PA+A^{T}P\right)  x+2v^{T}\left(  \frac{\partial h}{\partial
t}+\frac{\partial\sigma}{\partial t}+\frac{1}{\epsilon}\frac{\partial
h}{\partial u}\xi\right) \\
&  \text{ \ \ }-2v^{T}\left[  \frac{1}{\epsilon}\frac{\partial h}{\partial
u}-\left(  -K^{T}+\frac{\partial\sigma}{\partial x}\right)  B\right]
v+2x^{T}\left[  PB-A^{T}\left(  -K^{T}+\frac{\partial\sigma}{\partial
x}\right)  ^{T}\right]  v.
\end{align*}
By (\ref{PQ}), it follows $x^{T}\left(  PA+A^{T}P\right)  x\leq-\gamma
_{0}\left\Vert x\right\Vert ^{2},$ where $\gamma_{0}=\lambda_{\min}\left(
M\right)  .$ Then%
\begin{align}
\dot{V}  &  \leq-\gamma_{0}\left\Vert x\right\Vert ^{2}+2\left\Vert
v\right\Vert \left(  l_{h_{t}}\left\Vert u\right\Vert +l_{\sigma_{t}%
}\left\Vert x\right\Vert +d_{\sigma}+\frac{\overline{l}_{h_{u}}}{\epsilon
}\left\Vert \xi\right\Vert \right) \nonumber\\
&  \text{ \ \ }-2\left(  \frac{1}{\epsilon}\underline{l}_{h_{u}}-\frac{1}%
{2}\gamma_{1}\right)  \left\Vert v\right\Vert ^{2}+2\gamma_{2}\left\Vert
v\right\Vert \left\Vert x\right\Vert \text{\ \ \ (\textit{Assumptions 2-3} are
used)} \label{dv}%
\end{align}
Next, the relationship between $\left\Vert u\right\Vert $ and $\left\Vert
v\right\Vert $ needs to be derived to eliminate $\left\Vert u\right\Vert $ in
(\ref{dv}). By \textit{Lemma 1}, it follows%
\[
h\left(  t,u\right)  =h\left(  t,0\right)  +\int_{0}^{1}\left.  \frac{\partial
h}{\partial u}\right\vert _{x=su}ds\cdot u.
\]
Furthermore, since $v=h\left(  t,u\right)  -K^{T}x+\sigma\left(  t,x\right)
$, it follows%
\[
u=\left(  \int_{0}^{1}\left.  \frac{\partial h}{\partial u}\right\vert
_{x=su}ds\right)  ^{-1}\left(  v+K^{T}x-\sigma\left(  t,x\right)  -h\left(
t,0\right)  \right)  .
\]
Further by \textit{Assumptions 2-3},\textbf{\ }the equation above becomes%
\begin{align*}
\left\Vert u\right\Vert  &  \leq\frac{1}{\underline{l}_{h_{u}}}\left(
\left\Vert v\right\Vert +\left\Vert K\right\Vert \left\Vert x\right\Vert
+\left\Vert \sigma\left(  t,x\right)  \right\Vert \right) \\
&  \leq\frac{1}{\underline{l}_{h_{u}}}\left(  \left\Vert v\right\Vert +\left(
\left\Vert K\right\Vert +k_{\sigma}\right)  \left\Vert x\right\Vert
+\delta_{\sigma}\right)  .
\end{align*}
With the above inequality and the further inequality%
\[
2\left\Vert v\right\Vert \left(  \frac{l_{h_{t}}}{\underline{l}_{h_{u}}}%
\delta_{\sigma}+d_{\sigma}+\frac{\overline{l}_{h_{u}}}{\epsilon}\left\Vert
\xi\right\Vert \right)  \leq\frac{1}{\epsilon}\underline{l}_{h_{u}}v^{2}%
+\frac{\epsilon}{\underline{l}_{h_{u}}}\left(  \frac{l_{h_{t}}}{\underline
{l}_{h_{u}}}\delta_{\sigma}+d_{\sigma}+\frac{\overline{l}_{h_{u}}}{\epsilon
}\left\Vert \xi\right\Vert \right)  ^{2},
\]
the inequality (\ref{dv}) becomes%
\[
\dot{V}\leq-\gamma_{0}\left\Vert x\right\Vert ^{2}-\left(  \frac{1}{\epsilon
}\underline{l}_{h_{u}}-\gamma_{1}\right)  v^{2}+\frac{\epsilon}{\underline
{l}_{h_{u}}}\left(  \frac{l_{h_{t}}}{\underline{l}_{h_{u}}}\delta_{\sigma
}+d_{\sigma}+\frac{\overline{l}_{h_{u}}}{\epsilon}\left\Vert \xi\right\Vert
\right)  ^{2}+2\left(  \gamma_{2}+l_{\sigma_{t}}\right)  \left\Vert
v\right\Vert \left\Vert x\right\Vert .
\]
Since the inequality $2ab\leq a^{2}+b^{2}$ always holds, then%
\[
2\left(  \gamma_{2}+l_{\sigma_{t}}\right)  \left\Vert v\right\Vert \left\Vert
x\right\Vert \leq\frac{\gamma_{0}}{2}\left\Vert x\right\Vert ^{2}+\frac
{2}{\gamma_{0}}\left(  \gamma_{2}+l_{\sigma_{t}}\right)  ^{2}v^{2}.
\]
Consequently,%
\[
\dot{V}\leq-\frac{\gamma_{0}}{2}\left\Vert x\right\Vert ^{2}+\frac{\epsilon
}{\underline{l}_{h_{u}}}\left(  \frac{l_{h_{t}}}{\underline{l}_{h_{u}}}%
\delta_{\sigma}+d_{\sigma}+\frac{\overline{l}_{h_{u}}}{\epsilon}\left\Vert
\xi\right\Vert \right)  ^{2}-\left[  \frac{1}{\epsilon}\underline{l}_{h_{u}%
}-\gamma_{1}-\frac{2}{\gamma_{0}}\left(  \gamma_{2}+l_{\sigma_{t}}\right)
^{2}\right]  \left\Vert v\right\Vert ^{2}.
\]
Furthermore,%
\[
\dot{V}\leq-\eta\left(  \epsilon\right)  V+\frac{\epsilon}{\underline
{l}_{h_{u}}}\left(  \frac{l_{h_{t}}}{\underline{l}_{h_{u}}}\delta_{\sigma
}+d_{\sigma}+\frac{\overline{l}_{h_{u}}}{\epsilon}\left\Vert \xi\right\Vert
\right)  ^{2}%
\]
where $\eta\left(  \epsilon\right)  =\min(\frac{\gamma_{0}}{2\lambda_{\max
}\left(  P\right)  },\frac{1}{\epsilon}\underline{l}_{h_{u}}-\gamma_{1}%
-\frac{2}{\gamma_{0}}\left(  \gamma_{2}+l_{\sigma_{t}}\right)  ^{2}).$ If
(\ref{condition}) is satisfied, then $\eta\left(  \epsilon\right)  >0.$ So,
$V\left(  t\right)  \rightarrow\mathcal{B}\left(  \frac{1}{\eta\left(
\epsilon\right)  }\frac{\epsilon}{\underline{l}_{h_{u}}}\left(  \frac
{l_{h_{t}}}{\underline{l}_{h_{u}}}\delta_{\sigma}+d_{\sigma}+\frac
{\overline{l}_{h_{u}}}{\epsilon}\left\Vert \xi\right\Vert \right)
^{2}\right)  $ as $t\rightarrow\infty$, namely
\[
\left\Vert x\left(  t\right)  \right\Vert \rightarrow\mathcal{B}\left(
\sqrt{\frac{\epsilon}{\lambda_{\min}\left(  P\right)  \eta\left(
\epsilon\right)  \underline{l}_{h_{u}}}}\left(  \frac{l_{h_{t}}}{\underline
{l}_{h_{u}}}\delta_{\sigma}+d_{\sigma}+\frac{\overline{l}_{h_{u}}}{\epsilon
}\left\Vert \xi\right\Vert \right)  \right)
\]
as$\ t\rightarrow\infty,$ where $\mathcal{B}\left(  \delta\right)
\triangleq\left\{  \xi\in%
%TCIMACRO{\U{211d} }%
%BeginExpansion
\mathbb{R}
%EndExpansion
\left\vert \left\Vert \xi\right\Vert \leq\delta\right.  \right\}  ,$
$\delta>0.$ The notation $z\left(  t\right)  \rightarrow\mathcal{B}\left(
\delta\right)  $ means $\underset{y\in\mathcal{B}\left(  \delta\right)  }%
{\min}$ $\left\vert z\left(  t\right)  -y\right\vert \rightarrow0$ as
$t\rightarrow\infty.$ Since $\xi\rightarrow0$ as $t\rightarrow\infty,$
$\left\Vert x\left(  t\right)  \right\Vert \rightarrow\mathcal{B}\left(
\sqrt{\frac{\epsilon}{\lambda_{\min}\left(  P\right)  \eta\left(
\epsilon\right)  \underline{l}_{h_{u}}}}\left(  \frac{l_{h_{t}}}{\underline
{l}_{h_{u}}}\delta_{\sigma}+d_{\sigma}\right)  \right)  .$\ Furthermore, if
$l_{h_{t}}\delta_{\sigma}\left(  t\right)  \rightarrow0$ and$\ d_{\sigma
}\left(  t\right)  \rightarrow0,$ then the state $x\left(  t\right)
\rightarrow0$ as $t\rightarrow\infty.$

\end{document}